\documentclass[superscriptaddress, reprint, amsmath, amssymb, aps, pra, floatfix,nofootinbib]{revtex4-2}

\usepackage{bm}
\usepackage{algpseudocode}
\usepackage{algorithm}
\usepackage{booktabs}

\usepackage{mathtools}
\usepackage{amsfonts}
\usepackage{amssymb}
\usepackage{dsfont}
\usepackage{amsthm}

\usepackage[english]{babel}
\usepackage{CJKutf8}

\usepackage{graphicx}
\usepackage{amsmath}
\usepackage{lipsum}
\usepackage{wrapfig}
\usepackage{float}
\usepackage{enumitem} 
\usepackage{color}
\usepackage[normalem]{ulem}
\usepackage{url}
\usepackage{longtable}

\usepackage{comment}
\usepackage{array}

\def\bra#1{\ensuremath{\mathinner{\langle{#1}|}}}
\def\ket#1{\ensuremath{\mathinner{|{#1}\rangle}}}

\newcommand{\needcite}[1]{\textcolor{red}{[Ref needed]}}
\usepackage{qcircuit}

\usepackage{hyperref}
\hypersetup{colorlinks = true}

\begin{document}

\title{Loss tolerant cross-Kerr enhancement via modulated squeezing}

\author{Ankit Tiwari}
\affiliation{School of Electrical, Computer, and Energy Engineering, Arizona State University, Tempe, Arizona 85287, USA}

\author{Daniel Burgarth}
\affiliation{Department Physik, Friedrich-Alexander-Universit\"at Erlangen-N\"urnberg, Staudtstra\ss e 7, 91058 Erlangen, Germany}

\author{Linran Fan}
\affiliation{Chandra Department of Electrical and Computer Engineering, The University of Texas at Austin, Austin, TX, USA}

\author{Saikat Guha}
\affiliation{Department of Electrical and Computer Engineering, University of Maryland, 8228 Paint Branch Dr, College Park, MD 20742, USA}

\author{Christian Arenz}
\affiliation{School of Electrical, Computer, and Energy Engineering, Arizona State University, Tempe, Arizona 85287, USA}


\begin{abstract}
We develop squeezing protocols to enhance cross-Kerr interactions. We show that through alternating between squeezing along different quadratures of a single photonic mode, the cross-Kerr interaction strength can be generically amplified. As an application of the squeezing protocols, we discuss speeding up the deterministic implementation of controlled phase gates in photonic quantum computing architectures. We develop bounds that characterize how fast and strong single-mode squeezing has to be applied to achieve a desired gate error and show that the protocols can overcome photon losses. Finally, we discuss experimental realizations of the squeezing strategies in optical fibers and nanophotonic waveguides.    
\end{abstract}

\maketitle

\emph{Introduction ---}Photonic systems are among the most promising platforms for realizing a fault-tolerant quantum computer \cite{RevModPhys.79.135}. Qubits encoded in the presence of a single photon in one of two orthogonal modes, along with linear optical elements, such as phase shifters and beam splitters, and some non-linear process, enable universal quantum computing \cite{PhysRevA.52.3489, Milburn}. The non-linear process required to implement controlled phase gates on such {\em dual rail} photonic qubits is the cross-Kerr interaction \cite{PhysRevLett.62.2124}.

Unfortunately, the in-line Kerr effect for optical frequencies in non-linear material platforms used in photonic systems is typically extremely weak. Consequently, imparting a full $\pi$ phase shift for single-photon level signals~\cite{Shapiro2006} needed to realize a controlled phase gate is deterministically not achievable, which poses one of the main challenges for realizing a photonic quantum computer. The way the photonic quantum computing community has mitigated this shortcoming is by eliminating the in-line Kerr phase by the injection of ancilla single photons into an otherwise linear optical circuit~\cite{Pant2019}, leveraging an effect termed Boosted Bell State Measurement~\cite{Ewert2014}. Variants of this architecture underlie the platform of PsiQuantum, a company that is pursuing fault-tolerant photonic quantum computing~\cite{Bartolucci2021}. The shortcoming of this workaround is that despite the {\em boosting} of the success probability of two-qubit gates such as the controlled phase gate, the architecture still relies on probabilistic gates, hence massive amounts of multiplexed heralded operations, which drive the resource overheads to be astronomically high.

In this work, we address the problem of weak cross-Kerr interactions by developing protocols that amplify the cross-Kerr interaction through squeezing. This is achieved by utilizing a recently introduced protocol called \emph{Hamiltonian amplification} \cite{arenz2020amplification,burd2023experimental} in which the free cross-Kerr evolution is interspersed by single-mode squeezing transformations along different quadratures (i.e., the squeezing direction is modulated as a function of time). We show that Hamiltonian amplification can be used to generically enhance the cross-Kerr interaction strength between multiple optical modes of a quantum system independently of system and parameter details.

 \begin{figure}[t!]
 \centering
  \includegraphics[width=8cm]{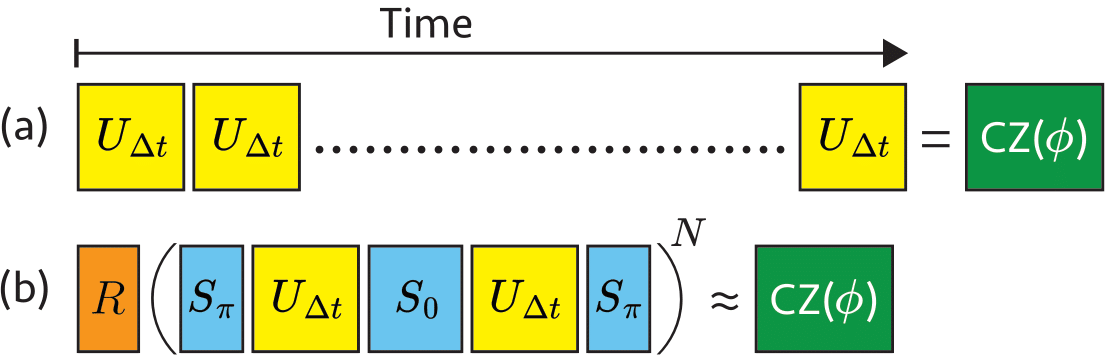}
 \caption{\label{fig:intro} Schematic representation of the squeezing protocols that are developed in this work. A controlled phase gate $\text{CZ}(\phi)$ (green) can be (deterministically) implemented by (a) repeating the evolution $U_{\Delta t}$ generated by the cross-Kerr interaction  (yellow) over a time interval $\Delta t$. Since the cross-Kerr interaction is typically weak, this process is unpractical as the time scale required to impart a full $\pi$ phase shift is beyond the coherence time of the system. Instead, in (b) the cross-Kerr evolution $U_{\Delta t}$ is interspersed with single-mode squeezing transformations $S_{\theta}$ (blue) along different quadratures ($\theta=0,\pi$) to enhance the cross-Kerr interaction. The wider blue squeezing box in the middle of the sequence indicates squeezing with twice the strength $r$ as $S^{\dagger}_{\pi}=S_{0}$, which is used to simplify the squeezing sequence given in \eqref{eq:HAsequence}. In this way only a few repetitions of $U_{\Delta t}$ and a phase shifter $R$ are required to implement $\text{CZ}(\pi)$. The resulting (Trotter) error is determined by the spacings $\Delta t=\frac{t}{2N}$ of $S_{\theta}$. }
\end{figure}

As depicted in Fig. \ref{fig:intro}, as an application of the modulated squeezing strategies, we consider speeding up the implementation of controlled phase gates in photonic quantum computing. We develop an error bound for the gate error that characterizes how fast squeezing has to be applied to achieve a desired speed-up. We go on to show that the developed squeezing sequences can outperform photon losses when sufficiently fast and strong squeezing is available. Finally, we discuss experimental implementations of the protocols in optical fibers and nanophotonic waveguides. We show that a significant improvement of currently observed controlled phase shifts is possible through the developed sequences.

\emph{Enhancing cross-Kerr interactions through Hamiltonian amplification ---}
The cross-Kerr interaction between two photonic modes described by quantum harmonic oscillators is described by the Hamiltonian 
\begin{align}
\label{eq:crossKerrH}
H=\chi a^{\dagger}ab^{\dagger}b,
\end{align}
where we set $\hbar=1$, $
\chi$ is the cross-Kerr interaction strength and $a,a^{\dagger}$ and $b,b^{\dagger}$ are the bosonic annihilation and creation operators of each mode.  

In general, several strategies exist that use squeezing to enhance interactions involving quantum harmonic oscillators \cite{QIN20241}. However, these strategies typically require fine tuning to the system's parameters and often rest on assumptions about the initial state of the system and the parameter regimes considered. For example, the strategy developed in \cite{bartkowiak2014quantum} to enhance cross-Kerr interactions uses single-mode squeezing and phase shifters that are precisely tuned to the cross-Kerr strength $\chi$ and requires a certain class of initial states. These requirements make it challenging to implement the protocol in the laboratory without introducing more noise and photon losses. A recently introduced approach called \emph{Hamiltonian amplification}
\cite{arenz2020amplification,burd2023experimental} can overcome these challenges. This strategy amplifies interactions by interspersing the free evolution with squeezing transformations along orthogonal quadratures to obtain an interaction enhancement independent of the initial state and parameter details. Below, we describe how cross-Kerr interactions can be enhanced through Hamiltonian amplification.

Consider modifying the dynamics $U_{\Delta t}=e^{-iH\Delta t }$ generated by the cross-Kerr interaction given by the Hamiltonian in Eq. \eqref{eq:crossKerrH} by alternating in time intervals $\Delta t=\frac{t}{2N}$ between squeezing a single mode along different quadratures described by the squeezing transformation $S_{\theta}=\exp\left[\frac{r}{2}\left(a^{2}e^{-i\theta}-a^{\dagger 2}e^{i\theta}\right)\right]$. Here, $\theta$ is the squeezing angle that describes
along which quadrature the mode is squeezed. We omit the explicit dependence of $S_{\theta}$ on the squeezing parameter $r$ that describes the squeezing strength. If the alternation is done sufficiently fast, described by the (Trotter) limit, 
\begin{align}\label{eq:HAsequence}
\lim_{N\to\infty}\left(S_{0}^{\dagger}U_{\Delta t}S_{0}S_{\pi}^{\dagger}U_{\Delta t}S_{\pi}\right)^{N}=e^{-iH_{{\lambda_{1}}}t},
\end{align}
the resulting dynamics is given by an evolution governed by the Hamiltonian $H_{\lambda_{1} } = \lambda_{1}H + H_{\nu_{1}}$ where  $\lambda_1=\cosh(2r)$ is the amplification factor of the cross-Kerr interaction and $H_{\nu_{1}} =  \chi \sinh^2(r)b^{\dagger}b$ causes a phase shift of the $b$ mode. We note that amplifying $H$ by $\lambda_1$ through the sequence \eqref{eq:HAsequence} does not require knowledge of the cross-Kerr strength $\chi$ and is independent of the initial state of the system.  By simultaneously squeezing both modes, larger amplification factors $\lambda_{2} = \cosh^2(2r)$ can be achieved, which we discuss next.  

Since in the limit of infinitely fast alternation $(N\to\infty)$ the resulting dynamics is governed by a map of the form $M(H)=\frac{1}{2}(S_{0}^{\dagger}HS_{0}+S_{\pi}^{\dagger}HS_{\pi})=H_{\lambda_{1}}$, composing two maps $M_{a}$ and $M_{b}$ that describe squeezing each mode $a$ and $b$, respectively, $M_{a}(M_{b}(H))=H_{\lambda_{2}}$, where $H_{\lambda_{2}} = \lambda_{2}H + H_{\nu_2}$ and $H_{\nu_2} =  \chi \cosh(2r)\sinh^2(r) (a^{\dagger}a + b^{\dagger}b) $ achieves amplification of the cross-Kerr interaction by a factor $\lambda_{2}=\cosh^{2}(2r)$.  The explicit form of the resulting squeezing sequence is given in the supplemental material. In general, composing maps of the form $M$ allows for amplifying generic multi-mode cross-Kerr interactions \cite{PhysRevResearch.6.023332,PhysRevA.110.022614} which we discuss in more detail in the supplemental material.

In order to investigate how fast squeezing along different quadratures needs to be applied to achieve the desired enhancement, we consider speeding up the implementation of a controlled phase gate in photonic quantum computing.

\emph{Speeding up the implementation of controlled phase gates in photonic quantum computing --- } The cross-Kerr interaction can be used to implement a controlled phase gate $\text{CZ}(\phi)=\text{diag}(1,1,1,e^{-i \phi})$ over dual rail qubits \cite{Nielsen_Chuang_2010, Ralph2011OpticalQC}. Indeed, the cross-Kerr evolution $U_{t}$ creates a controlled phase gate in the computational subspace $\mathcal C=\{\ket{00},\ket{10}, \ket{01}\ket{11}\}$, where $\ket{n}$, $n=0,1$, denote the eigenstates of $a^{\dagger}a$ and $b^{\dagger}b$. The controlled phase shift $\phi=\chi t$ is determined by the cross-Kerr interaction strength $\chi$ and the evolution time $t$. As such, in order to impart a desired phase shift, a sufficiently long coherent evolution is needed, depending on the cross-Kerr strength $\chi$. Thus, if $\chi$ is amplified through the sequence described in Eq. \eqref{eq:HAsequence}, a desired phase shift can be obtained for shorter evolution times. In particular, as depicted in Fig. \ref{fig:intro}, a controlled phase gate with an amplified phase shift $\phi=\chi t\lambda_{1}$ is implemented through the sequence \eqref{eq:HAsequence} for sufficiently larger $N$ followed by a phase shifter $R_{\nu_{1}}=e^{i H_{\nu_1}t}$.  The error that is introduced due to the Trotterization in Eq. \eqref{eq:HAsequence} is determined by the time interval $\Delta t$ of the Trotterization.

Using recently developed state-dependent Trotter bounds for infinite dimensional systems \cite{PhysRevA.107.L040201,vanLuijk_2024, PhysRevResearch.6.043155}, in the supplemental material we show that the gate error,  
\begin{align}\label{eq:error}
\epsilon_{N}(\phi)=\left\Vert \text{CZ}(\phi)-P\left[ R_{\nu_1 }\left(S_{0}^{\dagger}U_{\Delta t}S_{0}S_{\pi}^{\dagger}U_{\Delta t}S_{\pi}\right)^{N} \right]P \right\Vert_{\text{HS}}, 
\end{align}
for implementing a controlled phase shift $\phi=\chi t\lambda_{1}$ 
where $P$ is the projector into the computational subspace $\mathcal C$ and $\Vert \cdot \Vert_{\text{HS}}$ denotes the Hilbert-Schmidt norm is upper bounded by 
\begin{equation}\label{eq:bound}
\epsilon_{N}(\phi) \leq \frac{\chi^{2} t^{2}\cosh^{2}(2r)}{8N}f(r).
\end{equation}
For $r \gg 1$, the function $f(r)$ is approximately constant, such that $\epsilon=\mathcal O\left( \frac{\phi^{2}}{N}\right)$ where $\phi$ is the amplified phase shift. Consequently, for large squeezing strengths the Trotter error can be controlled by increasing the number of Trotter steps $N$,  \emph{independently} of how much the cross-Kerr interaction strength $\chi$ needs to be amplified to obtain the desired phase shift $\phi$.  The tightness of the bound in Eq. \eqref{eq:bound} is analyzed in more detail in the supplemental material.

\emph{Cross-Kerr enhancement in the presence of photon losses --- }We now turn to the question of how the amplification of the cross-Kerr interaction is influenced by photon losses.  Typically, photon losses are described by a quantum channel $\exp(\eta t (\mathcal  D_{a}+\mathcal  D_{b}))$ where
$\mathcal D_{L}(\cdot)=L (\cdot) L^{\dagger} + \frac{1}{2} \left( L^{\dagger}L(\cdot) +  (\cdot)L^{\dagger}L  \right )$ 
is a Lindblad operator that describes photon losses for each mode $a$ and $b$, and $\eta t$ is the overall loss rate. To gain some intuition for how amplification through the squeezing sequence described in Eq. \eqref{eq:HAsequence} can overcome photon losses, we start by assuming that photon losses occur between the squeezing transformations $S_{\theta}$ and the Trotterized cross-Kerr evolution $U_{\Delta t}$. In this case, photon losses can accumulate less compared to the case where squeezing is not present. This is because the amplification sequences can be shorter than the sequence that is formed by repeating the cross-Kerr evolution to obtain a desired phase shift, as schematically represented in Fig. \ref{fig:intro}. Consequently, in the amplified case the loss channel is less frequently applied, which results in less overall photon loss.

However, typically squeezing introduces more losses and also results in heating \cite{breuer2002theory}. In order to capture these effects, we proceed by considering a loss model in which photon losses also occur during the cross-Kerr evolution. With further details found in the supplemental material, this is achieved by weakly coupling each mode via a beam-splitter interaction to auxiliary modes that are assumed to be in the vacuum state \cite{PhysRevA.89.042309}. After each time interval $\Delta t$ the auxiliary modes are traced out to obtain a quantum channel $\mathcal E_{\Delta t}^{(j)}$ whose explicit form is given in the supplemental material. Here, $j$ labels the squeezing transformation that is applied before and after $\Delta t$. We show that when squeezing is applied to both modes, for sufficiently many Trotter steps $N$ the resulting sequence $(\prod_{j} \mathcal E_{\Delta t}^{(j)})^{N}$ is given by the quantum channel 
\begin{align}
\label{eq:lossmodelL}
\mathcal E_{t}=\exp([\mathcal{H}_{\lambda_2}+\cosh^2(r)\mathcal L_{l}+\sinh^2(r)\mathcal L_{h}] t)
\end{align}
where $\mathcal{H}_{\lambda_{2}}=-i[H_{\lambda_{2}},\cdot]$ is the super operator describing the cross-Kerr interaction, $\mathcal L_{l}=\eta(\mathcal D_{a}+\mathcal D_{b})$ describes the amplified photon losses and $\mathcal L_{h}=\eta (\mathcal D_{a^{\dagger}}+\mathcal D_{b^{\dagger}})$ are newly introduced heating processes. We note that for $r=0$, i.e., when no squeezing is present, the resulting quantum channel $\mathcal E_{t}=\exp([\mathcal H+\mathcal L_{l}]t)$, where $\mathcal{H} = -i[H,\cdot]$, describes photon losses that occur during the cross-Kerr evolution. We further remark that directly modeling photon losses during the cross-Kerr evolution through $\mathcal L_{l}$ yields for sufficiently large $N$ the same quantum channel as shown in Eq. \eqref{eq:lossmodelL}. A similar loss model was also used in \cite{bartkowiak2014quantum} to investigate squeezing-enhanced cross-Kerr interactions in the presence of photon losses, finding that photon losses can significantly impact the performance of the protocol in \cite{bartkowiak2014quantum}. Here we show that when squeezing can be applied to both modes,  the corresponding Hamiltonian amplification sequence resulting in the loss channel given in Eq. \eqref{eq:lossmodelL} can overcome photon losses, which we discuss next.

\emph{Overcoming photon losses --- }When squeezing is applied to both modes, the cross-Kerr interaction is amplified by $\cosh^{2}(2r)$ while losses and heating process occur with rates $\propto \cosh^{2}(r)$ and $\propto \sinh^2(r)$, respectively. We observe that the cross-Kerr interaction is more enhanced than the loss and heating processes, which can be traced back to the linear coupling of each mode via the beam splitter interaction to the auxiliary modes. As such, a full $\pi$ phase shift is obtained at a shorter time $t=\frac{\pi}{\chi \cosh^{2}(2r)}$ at which photon losses and heating processes occur with smaller rates $\approx e^{-2r}\frac{\eta}{\chi}$. As this observation is independent of the loss rate $\eta$, we refer to this behavior as ``loss-tolerant''.  

\begin{figure}[htbp]
\centering
\includegraphics[scale=1, width=8cm]{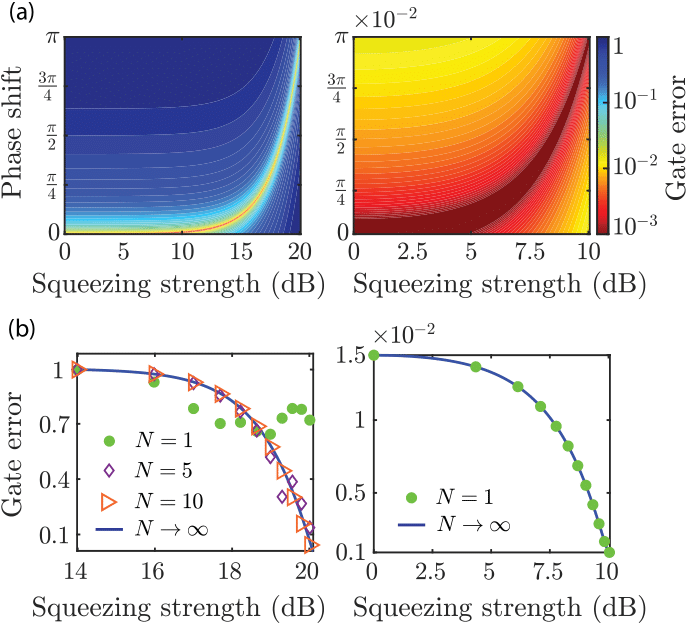}
\caption{\label{fig:colormaps and Trotterized channel} Gate error for implementing a controlled phase gate in the presence of photon losses when the cross-Kerr interaction is amplified through single-mode squeezing applied to both modes. The bare phase shift $\chi t = 1.2\times 10^{-3}$ and loss rate $\eta t = 5.76\times10^{-4}\,\text{dB}$ was chosen according to the parameters reported in optical fibers \cite{venkataraman2013phase, amrani2021low}. The color maps in (a) show the gate error as a function of the squeezing strength $r$ for phase shifts up to $\phi=\pi$ (left panel) and $\phi=\frac{\pi}{100}$ (right panel) based on the quantum channel in  Eq. \eqref{eq:lossmodelL} that is obtained in the limit $N\to \infty$.  In (b) the gate error is plotted for a fixed phase shift $\phi=\pi$ (left) and $\phi=\frac{\pi}{100}$ (right) for different Trotter steps $N=1$ (green circles), $N=5$ (purple diamonds) and $N=10$ (orange triangles). In all cases, the gate error was normalized to $1$ by dividing by its largest value.} 
\end{figure}

In Fig. \ref{fig:colormaps and Trotterized channel} we investigate the loss model given by the quantum channel $\mathcal E_{t}$ in Eq. \eqref{eq:lossmodelL}. For the numerical simulations we chose $\chi t = 1.2\times 10^{-3}$ and $\eta t=  5.76\times10^{-4}\,\text{dB}$, which correspond to the phase shift and loss rates reported for optical (photonic crystal) fibers in \cite{venkataraman2013phase} and \cite{amrani2021low}, respectively. Throughout this work, we express phase shifts in radians. 

In Fig. \ref{fig:colormaps and Trotterized channel} (a) we plot the gate error $\Vert \mathcal{CZ}_{\phi} - \mathcal{P}(\mathcal R_{\nu_2}\mathcal E_{t})\mathcal{P}\Vert_{\text{HS}}$ as a function of the squeezing strength $r$ for different phase shifts, i.e., up to a full $\pi$ phase shift (left panel) and up to a $\frac{\pi}{100}$ phase shift (right panel). Here, $\mathcal{CZ}_{\phi}(\cdot) = \text{CZ}(\phi)\,(\cdot)\,\text{CZ}^{\dagger}(\phi)$ and $\mathcal{R}_{\nu_2} = R_{\nu_2} (\cdot)R^{\dagger}_{\nu_2}$ are the quantum channels that describe the controlled phase gate and the phase shifter $R_{\nu_2} = e^{i H_{\nu_2} t}$, respectively. The superoperator $\mathcal{P} = P (\cdot)P$ describes the projection into the computational subspace. In the numerical simulations, we used the matrix representation of the quantum channels and superoperators obtained from row vectorization \cite{am2015three} of the density operator. 

The color map plots in Fig. \ref{fig:colormaps and Trotterized channel} (a) suggest that without amplification only phase shifts $\phi \leq \frac{\pi}{100}$ can be achieved with gate errors $\epsilon \leq 10^{-2}$. The situation changes when the cross-Kerr interaction is amplified by applying squeezing to both modes. A full $\pi$ phase shift with $\epsilon \approx 10^{-2}$ can be obtained when enough single-mode squeezing $r\approx 20\,\text{dB}$ is available. The colormap plot shown in the right panel in Fig. \ref{fig:colormaps and Trotterized channel} (a) shows the phase shift improvements that are possible in existing platforms. We observe that with a squeezing strength of $r \approx 8\,\text{dB}$, achievable in optical fibers \cite{liu2024entanglement} and nanophotonic waveguides \cite{kashiwazaki2023over}, about an order of magnitude improvement in the gate error to create a $\frac{\pi}{100}$ phase shift is possible.

In Fig. \ref{fig:colormaps and Trotterized channel} (b) we study the effect of Trotterization for a fixed phase shift $\phi=\pi$ (left panel) and $\phi=\frac{\pi}{100}$ (right panel). Fig \ref{fig:colormaps and Trotterized channel} (b) shows the gate error $\Vert \mathcal{CZ}_{\phi} - \mathcal{P}(\mathcal{R}_{\nu_{2}}  (\prod_{j} \mathcal E_{\Delta t}^{(j)})^{N})\mathcal{P}\Vert_{\text{HS}}$ for $N = 1$ (green circles), $N = 5$  (purple diamonds) and $N=10$  (orange triangles) Trotter steps. For comparison, we show the gate error determined for $N\to \infty$ by the quantum channel $\mathcal E_{t}$ in Eq. \eqref{eq:lossmodelL} as a blue line. We see that already for $N=1$ Trotter steps the behavior of the gate error is remarkably well described by the channel $\mathcal E_{t}$ when small phase shifts are considered (right panel). However, from the left panel in Fig. \ref{fig:colormaps and Trotterized channel} we see that to obtain a full $\pi$ phase shift, more Trotter steps $N$ are needed to achieve the gate errors determined by $\mathcal E_{t}$. For more details regarding the behavior of the evolution as a function of $N$ we refer to the supplemental material. There, we also study imperfections in the squeezing angle. We show that the squeezing sequence that led to the curve shown in the right panel of Fig. \ref{fig:colormaps and Trotterized channel} (b)  can tolerate a moderate amount of noise in the squeezing angle, which depends on how much the two modes are squeezed.

\emph{Potential experimental implementations --- }Experimental platforms in which cross-Kerr modulated phase shifts can be observed while photon losses remain low and sufficiently strong single-mode squeezing is available are ideal candidates for implementing the developed squeezing sequences. 
These requirements are for example met in optical fibers \cite{PhysRevLett.66.153,Debord:17,Bergman:91, Potasek:87, PhysRevA.88.013819,matsuda2009observation,Inoue:09, venkataraman2013phase, liu2024entanglement} and nanophotonic waveguides \cite{kashiwazaki2023over, kashiwazaki2020continuous, Pysher:09,Chen:22, Zhang:16, Hsieh:07, cui2022situ,Chang:17, Liu:22} where the spacing of the squeezers (i.e., in the fiber or waveguide) determines the time interval $\Delta t$ of the Trotterization.

In particular, the parameters used in the numerical simulations shown in Fig. \ref{fig:colormaps and Trotterized channel} were taken from an experimental setting that utilizes a rubidium vapour-filled hollow-core photonic bandgap fiber \cite{venkataraman2013phase}. In this platform, a cross-Kerr modulated phase shift of $3\times 10^{-4}$ is produced over
a $\approx 9 \text{ cm}$ long fiber. Assuming that this phase shift can be replicated four times within a single Trotter step in the low-loss optical fiber described in \cite{amrani2021low}, we estimated the total loss rate used in the numerical simulations as $\eta t = 5.76\times10^{-4}\,\text{dB}$ per mode. The overall loss rate is computed through the minimum attenuation rate $ = 1.6\times10^{-5}\,\text{dB/cm}$ reported in \cite{amrani2021low} for a hollow-core photonic crystal fiber, assuming a fiber with a total length of $36\,\text{cm}$. We further note that the attenuation rate reported in \cite{amrani2021low} has been demonstrated for a field at $1050\text{ nm}$ wavelength, whereas the cross-Kerr modulated phase shift in \cite{venkataraman2013phase} is observed in a field at $\approx 780\text{ nm}$. In a recent work \cite{liu2024entanglement} a squeezing strength of $r =  7.5\,\text{dB}$ has been generated in optical fibers type systems, which would within the considered parameter regime allow for improving the phase shifts currently achievable in these platforms by about an order of magnitude with a gate error of $\epsilon\leq 10^{-2}$ (see Fig. \ref{fig:colormaps and Trotterized channel} (b)). As such, we expect that significantly larger controlled phase shifts may be achievable in optical fibers through the developed squeezing protocols.

Nanophotonic waveguides are also a viable platform for implementing the cross-Kerr amplification sequences. Both, squeezed light \cite{kashiwazaki2020continuous, Pysher:09,Chen:22} and cross-Kerr modulated phase shifts \cite{Zhang:16, Hsieh:07, cui2022situ} can be generated in these systems. Squeezed light with a squeezing strength as large as $r=8\,\text{dB}$ has been generated in nanophotonic waveguides \cite{kashiwazaki2023over}, which corresponds to an amplification factor of $\cosh^{2}(2r) \approx 10.45$. Meanwhile, cross-Kerr modulated phase shifts of the order of $10^{-7}$ are expected in nanophotonic waveguides with waveguide losses  $\approx2\,\text{dB/cm}$ \cite{10.1063/1.3257378}. 
In recent studies a significant reduction in photon losses has been demonstrated in lithium niobate and silicon nitride waveguides, achieving loss rates of $4 \times 10^{-2}\,\text{dB/cm}$ \cite{li2023high} and $3.4 \times 10^{-4}\,\text{dB/cm}$ \cite{Liu:22}, respectively. We expect that further improvements in hybrid systems \cite{churaev2023heterogeneously,Chang:17} will pave the way for achieving large controlled phase shifts in nanophotonic waveguides through the protocols developed here.

\emph{Conclusions ---}We have developed single-mode squeezing protocols that enhance cross-Kerr interactions. In particular, we showed that the multi-mode cross-Kerr interaction strength can be amplified by alternating between squeezing along orthogonal quadratures of a single photonic mode.  
We found that the interaction enhancement scales exponentially in the number of modes involved. This observation may allow for overcoming challenges in systems where only small single-mode squeezing is available, but squeezing can simultaneously be applied to all modes that are coupled via multi-mode cross-Kerr interactions.

As a key application of the developed squeezing protocols, we considered the deterministic implementation of controlled phase gates in photonic quantum computing. We showed that the hard-to-realize strong cross-Kerr interaction at single-photon levels needed for realizing a controlled phase gate could be realized in a photon loss-tolerant manner. We also discussed concrete realizations of the proposed concept in optical fibers and nanophotonic waveguides, which suggests that the implementation of the scheme is feasible with current technology.

In future work, it would be instructive to perform a thorough resource evaluation to realize fault-tolerant quantum computing over dual-rail photonic qubits using the developed protocols, assuming realistic devices, and compare that with the conventional approach that rests on multiplexed probabilistic resource-state preparation and fusion-based quantum computations~\cite{Bartolucci2021}. Furthermore, the principles in this work could be leveraged to amplify other non-Gaussian Hamiltonians such as the cubic-phase Hamiltonian, which has applications in GKP-qubit photonic quantum computing~\cite{PhysRevA.64.012310}, more efficiently compared with the alternative method of probabilistic heralded-multiplexed preparation of cubic-phase ancillas~\cite{PhysRevA.100.052301}. Our work could also find applications broadly in the universal bosonic circuit synthesis problem, which could help realize quantum optimal measurements for applications such as optical codeword discrimination for attaining the Holevo limit of optical communications capacity~\cite{PhysRevLett.106.240502,6302198}, and designing non-Gaussian receivers for distributed photonic sensors~\cite{PhysRevResearch.3.033114,PhysRevResearch.3.033011}.

\emph{Acknowledgements ---} AT and SG thank Chaohan Cui for helpful discussions. SG, CA, and AT acknowledge funding support from the Air Force Office of Scientific Research (AFOSR) under the award FA9550-24-1-0139. CA and AT acknowledge support from the Quantum Collaborative and the SenSip Center at Arizona State University. 

\bibliography{References.bib}

\onecolumngrid

\newpage

\begin{center}
\textbf{\large Supplemental Material} 
\end{center}
\appendix
\renewcommand{\appendixname}{}

In this supplemental material, we give the explicit form of the squeezing sequence when squeezing is applied to both modes, discuss enhancing multi-mode cross-Kerr interactions, derive the bound given in Eq. (4) in the main body of the manuscript and investigate its tightness. We also give more details for the photon loss model considered in the manuscript and study the performance of the squeezing sequences when errors in the squeezing angles are present. 

\section{Single-mode squeezing applied to two modes}

By applying single-mode squeezing transformations $S_{x,\theta}$ to both modes, where $x\in \{a,b\}$ indicates which mode is squeezed, the cross-Kerr interaction can be amplified by a factor of  $\cosh^{2}(2r)$. The corresponding squeezing 
sequence reads
\begin{equation}\label{eq:two mode HA sequence}
\begin{aligned}
&\lim_{N\to\infty}\left(S_{a,0}^{\dagger}S_{b,0}^{\dagger}U_{\Delta t}S_{a,0}S_{b,0}S_{a,\pi}^{\dagger}S_{b,0}^{\dagger}U_{\Delta t}S_{a,\pi}S_{b,0}S_{a,0}^{\dagger}S_{b,\pi}^{\dagger}U_{\Delta t}S_{a,0}S_{b,\pi}S_{a,\pi}^{\dagger}S_{b,\pi}^{\dagger}U_{\Delta t}S_{a,\pi}S_{b,\pi}\right)^{N}
\\
&=e^{-iH_{\lambda_2}t},\color{black}
\end{aligned}
\end{equation}
where $H_{\lambda_2} =  \cosh^2(2r)H +  \chi \cosh(2r)\sinh^2(r) (a^{\dagger}a + b^{\dagger}b)$ and  $\Delta t = \frac{t}{4N}$.

\section{Enhancing multi-mode cross-Kerr interactions}
In this section, we discuss Hamiltonian amplification of a multi-mode cross-Kerr interactions. We consider the cross-Kerr interaction
\begin{align}
\label{eq:multimodecrossKerr}
H=\chi_{n} \prod_{j=1}^{n}a^{\dagger}_{j}a_{j},
\end{align}
between $n$ modes described by the bosonic annihilation and creation operators $a_{j}$ and $a_{j}^{\dagger}$. If we denote by $M_{j}$ the map that describes amplifying the term $a_{j}^{\dagger}a_{j}$ by a factor $\cosh(2r)$ through applying  squeezing to the $j$th mode, then the composition, 
\begin{align}
\prod_{j=1}^{n}M_{j}(H)=H_{\lambda_{n}},
\end{align}
where the amplified Hamiltonian given by
\begin{equation}\label{eq: multi mode amplified H}
\begin{aligned}
H_{\lambda_{n}} &= \cosh^n(2r) H  + H_{\nu_n},\text{ where }\\
H_{\nu_{n}} &= \chi_{n} \left(\sum^{n-1}_{k = 1}  \cosh^{n-k}(2r) \sinh^{2k}(r) \sum_{\substack{I \subset \{1,\hdots,n \} \\ \lvert I \rvert = n -k   } } \prod_{j\in I}a^{\dagger}_{j}a_{j} \right),
\end{aligned}
\end{equation}
allows for amplifying the multi-mode cross-Kerr interaction \eqref{eq:multimodecrossKerr} by a factor $\cosh^{n}(2r)\approx \cosh(2nr)$ (for $r\gg 1$). Thus, the amplified multi-mode cross-Kerr strength increases exponentially in the number of modes $n$ involved, while the single-mode squeezing strength $r$ can remain fixed. The additional terms $H_{\nu_{1}} $describe cross-Kerr interactions between $k$ modes where $k<n$ (when $n>2$) and terms $a^{\dagger}_{j}a_{j}$ that give rise to phase shifts.  
\\
\\
We note that the loss and heating processes obtained in the implementation of a controlled phase gate in the presence of losses can be outperformed even more when $n$-mode cross-Kerr interactions described in  \eqref{eq:multimodecrossKerr} are used to implement the controlled phase gates. If we assume that each mode couples linearly via a beam splitter interaction to auxiliary vacuum modes, as we discussed in detail in section \ref{sec: photon loss model}, the time to implement a full $\pi$ phase shift is then reduced to $t=\frac{\pi}{\chi \cosh^{n}(2r)}$ while at $t$ loss and heating processes occur with rates $\approx e^{-2nr}\frac{\eta}{\chi}$. We remark that in order to implement a controlled phase gate through the multi-mode interaction in \eqref{eq: multi mode amplified H}, resources that include cross-Kerr interactions between fewer modes are required to implement the transformation $R_{\nu_n} = e^{i H_{\nu_n} t}$ to cancel the extra term $H_{\nu_{n}}$ in \eqref{eq: multi mode amplified H}. For the two mode case discussed in the main body of the manuscript, the transformation $R_{\nu_{2}}$ can simply be implemented through phase shifters. 

\section{Bounding the Trotter error}\label{sec:TrotterError}
Here we derive the upper bound for the gate error $\epsilon_{N}(\phi)$ given in Eq. (6) in the main body of the manuscript using the state-dependent
Trotter bounds for unbounded Hamiltonians from \cite{PhysRevResearch.6.043155}. In particular, we will make use of the Theorem: 
\\

\textbf{Theorem 1}: \textit{Let $H_{1}$ be self-adjoint on $D(H_{1})$ and $H_{2}$ be
self-adjoint on $D(H_{2})$. Let $\ket{\varphi}$ be an eigenstate of $H=H_{1} +H_{2}$
with eigenvalue $h$, i.e., $H \ket{\varphi}$ = $h \ket{\varphi}$. If $ \ket{\varphi}$ $\in$ $D(H^{2}_{1})\cap D(H^{2}_{2})$, then}
\begin{equation} \label{Theo:Theorem}   
\begin{aligned}
 \epsilon_{N}({t;\ket{\varphi}}) \leq \frac{t^{2}}{2N} \biggl( \left\Vert (H_{1} - g  )^{2} \ket{\varphi} \right\Vert + \left\Vert ( H_{2} - h + g)^{2} \ket{\varphi} \right\Vert \biggr),
\end{aligned}
\end{equation}
\textit{for all t, g $\in \mathbb{R}$, and the Trotter product formulae converges on $\ket{\varphi}$.}
\\
\\
Here, the Trotter error $\epsilon_{N}({t;\ket{\varphi}})$ is given by 
\begin{equation}\label{eq:Theorem error eucl. norm}
\epsilon_{N}(t;\ket{\varphi}) = \left\Vert \biggl(e^{-iH t} -  \left( e^{-iH_{1}\frac{t}{N}} e^{-iH_{2}\frac{t}{N}}\right) ^{N} \biggr) \ket{\varphi} \right\Vert,    
\end{equation}
where $\Vert \cdot \Vert$ denotes the Euclidean vector norm.
\\
\\
Before we apply Theorem 1, we first show that the domain assumptions are satisfied. We note that $R_{\nu_{1}}\left(S_{0}^{\dagger}U_{\Delta t}S_{0}S_{\pi}^{\dagger}U_{\Delta t}S_{\pi}\right)^{N} = \left(e^{-iH_{1}\Delta t}e^{-iH_{2}\Delta t}\right)^{N}$, where
\begin{equation}
\begin{aligned}
H_{1}&=\chi \biggl( \cosh (2r) a^{\dagger}a - \frac{\sinh (2r)}{2} (a^{\dagger}a^{\dagger} + aa ) \biggr) b^{\dagger}b, \\
H_{2}&= \chi \biggl( \cosh (2r) a^{\dagger}a + \frac{\sinh (2r)}{2} (a^{\dagger}a^{\dagger} + aa) \biggr) b^{\dagger}b,
\end{aligned}
\end{equation}
and define the space of finite excitations by 
\[
\mathcal{\mathcal{F}}=\left\{ |\varphi\rangle\in L^{2}(\mathbb{R}^{2})\mid\exists N,M\in\mathbb{N}:|\varphi\rangle=\sum_{k=0}^{N}\sum_{\ell=0}^{M}c_{k\ell}|k\ell\rangle\right\}, 
\]
where $\ket{lk}$ are the fock states of mode $a$ and mode $b$. We first show that $\mathcal F$ is a set of analytic vectors of the Hamiltonians, 
\begin{equation}\label{eq:classofH}
H=\left(\alpha a^{\dagger}a+\beta(a^{\dagger}a^{\dagger}+aa)\right)b^{\dagger}b
\end{equation}
with $\alpha,\beta\in\mathbb{R}$, noting that $H_{1}$ and $H_{2}$ defined above are of this form. 
\\
\\
We use the definition of analytic vectors $\ket{\varphi}$ in \cite{reed2003methods} (Section X.6, p201), 
\[
\sum_{n=0}^{\infty}\frac{\left\Vert H^{n}|\varphi\rangle\right\Vert }{n!}t^{n}<\infty
\]
for some $t>0$. Let $\gamma=\max\left\{ |\alpha|,|\beta|\right\} $
and for a fixed $|\varphi\rangle\in\mathcal{F}$ and let $K=\max\left\{ N,M\right\}$. 
Then
\[
H^{n}|k\ell\rangle=\left(\alpha a^{\dagger}a+\beta(a^{\dagger}a^{\dagger}+aa)\right)^{n}\ell^{n}|k\ell\rangle
\]
and 
\[
\left\Vert H^{n}|\varphi\rangle\right\Vert \le\gamma^{n}3^{n}\sqrt{\frac{(K+2n)!}{K!}}K^{n}
\]
where we upper bounded the expansion of operators by the worst case
where only strings of creation operators with $\left(a^{\dagger}\right)^{2n}|K\rangle=\sqrt{\frac{(K+2n)!}{K!}}|K+2n\rangle$
on the first mode appear, and upper bounded the contributions from
$|\varphi\rangle$ by its highest number $K$ of photons. Now let 
\[
c_{n}=\frac{(3\gamma Kt)^{n}\sqrt{\frac{(K+2n)!}{K!}}}{n!},
\]
then
\[
\frac{c_{n+1}}{c_{n}}=\frac{(3\gamma Kt)\sqrt{(K+2n+2)(K+2n+1)}}{(n+1)}\rightarrow6\gamma Kt
\]
and by the ratio test the series converges for $t<\frac{1}{6\gamma K}.$ Consequently, any $\ket{\varphi}\in \mathcal F$ is an analytic vector of $H$ given in \eqref{eq:classofH}.   
\\
\\
We note that $H$ being essentially self-adjoint on $\mathcal{F}$ follows from Corollary 2 in \cite{reed2003methods} (Section X.6, p202), where $H\mathcal{F}=\mathcal{F}$ follows from the fact that $H$ maximally creates two new photons and the denseness of $\mathcal{F}$ is used. In abuse
of notation, we denote the closure of $H$ on $\mathcal{F}$ also as
$H$. By definition, for some self-adjoint $H$ the operator $H^{2}$ is the self-adjoint
operator with domain $D(H^{2})=\left\{ |\psi\rangle\in D(H)\mid H|\psi\rangle\in D(H)\right\} .$
Since $\mathcal{F}\in D(H)$ and $H\mathcal{F}\in D(H),$ any $|\varphi\rangle\in\mathcal{F}$
is in $D(H^{2}).$ 
\\
\\
In order to apply Theorem 1 to upper bound the gate error $\epsilon_{N}(\phi)$, we first notice that 
\begin{align}
\epsilon_{N}(\phi) &=  \left \Vert P \left( \text{CZ}(\phi)-\left(e^{-iH_{1}\Delta t}e^{-iH_{2}\Delta t}\right)^{N} \right)P \right \Vert_{\text{HS}} \nonumber \\
& \leq \sqrt{\sum_{x\in \{0,1\}^{2}}\left\Vert\left(  \text{CZ}(\phi)-(e^{-iH_{1}\frac{t}{2N}}e^{-iH_{2}\frac{t}{2N}})^{N}\right)\ket{x} \right \Vert ^{2}}
\end{align}
where we used that for any operator $A$  we have, 
\begin{equation}
\begin{aligned}
     \left \Vert PAP \right \Vert_{\text{HS}} =& \sqrt{\text{Tr}(PA^{\dagger}P AP)}\\
     =& \sqrt{\sum_{x\in \{0,1\}^{2} }\bra{x}A^{\dagger}PA\ket{x}}\\
     \leq &\sqrt{\sum_{ x\in \{0,1\}^{2}}\left \Vert A \ket{x} \right \Vert^{2} }.
\end{aligned}
\end{equation}
We now can use Theorem 1 to further upper bound the right-hand side to obtain 
\begin{align}
\epsilon_{N}(\phi)\leq \sqrt{\sum_{x\in \{0,1\}^{2}}\left(\Vert H_{1}^{2}\ket{x} \Vert +\Vert (H_{2}-h_{x})^{2}\ket{x} \Vert \right)^{2}}, 
\end{align}
where we picked $g=0$ in \eqref{Theo:Theorem} and $h_{x}$ is the eigenvalue of $H$ that corresponds to the eigenstate $\ket{x}$. We note that $h_{x}=0$ for $x\in\{00,01,10\}$ and $h_{11}=2\chi \cosh(2r)$. Since $\Vert H_{1}^{2}\ket{00}\Vert=\Vert H_{1}^{2}\ket{10}\Vert=\Vert H_{2}^{2}\ket{00}\Vert=\Vert H_{2}^{2}\ket{10}\Vert=0$ we further find that
\begin{align}
\epsilon_{N}(\phi)\leq \sqrt{\left(\Vert H_{1}^{2}\ket{01} \Vert +\Vert H_{2}^{2}\ket{01} \Vert \right)^{2}+\left(\Vert H_{1}^{2}\ket{11} \Vert +\Vert (H_{2}-h_{11})^{2}\ket{11} \Vert \right)^{2}}. 
\end{align}
As 
\begin{align}
\left \Vert H_{1}^{2} \ket{01} \right \Vert = \biggl|\!\biggl|&  \frac{ \chi^{2} \sinh^{2}(2r)}{2}\ket{01}-\sqrt{2}\chi^{2} \sinh(2r)\cosh(2r) \ket{21} +\frac{\sqrt{6}\chi^{2} \sinh^{2}(2r)}{2}\ket{41}\biggl|\!\biggl|, \\
\left \Vert H_{2}^{2} \ket{01} \right \Vert = \biggl|\!\biggl|&  \frac{ \chi^{2} \sinh^{2}(2r)}{2}\ket{01}+\sqrt{2}\chi^{2} \sinh(2r)\cosh(2r) \ket{21} +\frac{\sqrt{6}\chi^{2} \sinh^{2}(2r)}{2}\ket{41}\biggl|\!\biggl|,
\end{align}
and
\begin{equation}
\begin{aligned}
\left \Vert H_{1}^{2} \ket{11}\right\Vert = \biggl|\!\biggl|&\Biggl(\chi^{2}\cosh^{2}(2r) 
+ \frac{3\chi^{2}\sinh^{2}{(2r)}}{2} \Biggr)\ket{11}-2\sqrt{6}\chi^{2}\sinh(2r)\cosh(2r)\ket{31} \\&+ \frac{\sqrt{30}\chi^{2}\sinh^{2}(2r)}{2}\ket{51}\biggl|\!\biggl|, \\
\end{aligned}
\end{equation}
\begin{equation}
\begin{aligned}
\left \Vert (H_{2}-h_{11})^{2} \ket{11} \right \Vert = \biggl|\!\biggl|&\Biggl(\chi^{2}\cosh^{2}(2r) - 2h_{11} \chi \cosh(2r) 
+ \frac{3\chi^{2}\sinh^{2}{(2r)}}{2}  + h_{11}^{2} \Biggr)\ket{11}
\\
& + \Bigl(2\sqrt{6}\chi^{2}\sinh(2r)\cosh(2r)-\sqrt{6}h_{11} \chi \sinh(2r)\Bigr)\ket{31} \\ &+ \frac{\sqrt{30}\chi^{2}\sinh^{2}(2r)}{2}\ket{51}\biggl|\!\biggl|, 
\end{aligned}
\end{equation}
we arrive at the final result
\begin{align}\label{eq: upper bound final result}
\epsilon_{N}(\phi) \leq &\frac{\chi^{2} t^{2}\cosh^{2}(2r)}{8N} f(r), 
\end{align}
where 
\begin{align}\label{eq: f(r) in bounds}
 f(r) = \biggl[ &7\tanh^{4}(2r) +  8 \tanh^{2}(2r)  + \frac{1}{4} \biggl(  \bigl( 4 + 108 \tanh^{2}{(2r)} + 39 \tanh^{4}(2r)  \bigr)^{\frac{1}{2}} + \bigl( 4 + 12\tanh^{2}{(2r)}\\  & + 39\tanh^{4}{(2r)} \bigr)^{\frac{1}{2}} \biggr)^{2} \biggr]^{\frac{1}{2}}.\notag   
\end{align}

 \begin{figure}[ht]
 \centering
  \includegraphics[scale=0.75]{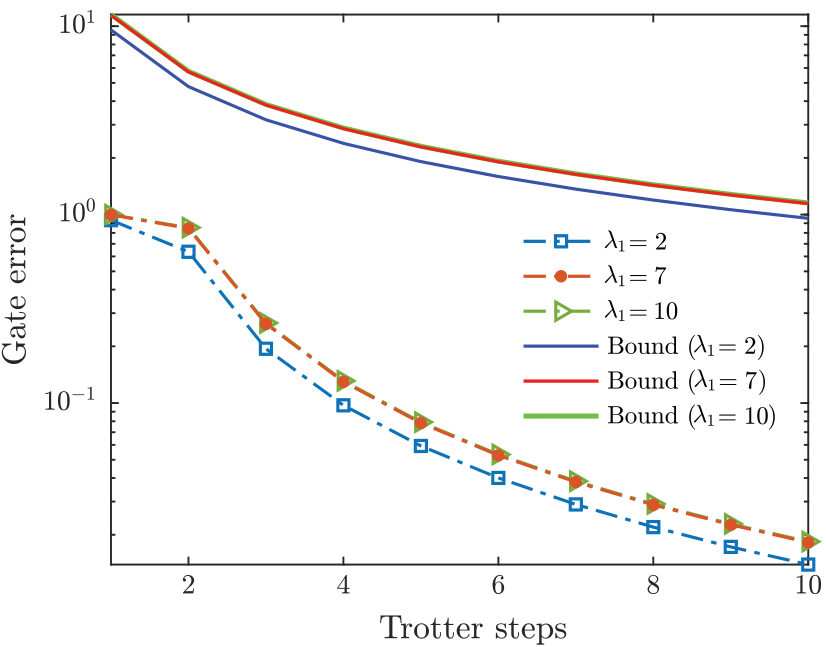}
 \caption{\label{fig:upperbounds}Gate error \eqref{eq: upper bound final result} shown as dashed lines for implementing controlled phase gate $\text{CZ}(\pi)$ 
 as a function of the number of Trotter steps $N$ for different values of the cross-Kerr phase shift $\chi t= \frac{\pi}{2}, \frac{\pi}{7}, \frac{\pi}{10}$ for different amplification factors $\lambda_1 =2,7,10$. The corresponding upper bounds given by \eqref{eq: f(r) in bounds} are shown as solid lines. The gate errors were normalized to 1 by dividing by its largest value. For comparison, the upper bounds were also normalized by dividing by the corresponding largest gate error.}
\end{figure}

Next, we analyze the tightness of the upper bound \eqref{eq: upper bound final result}. In Fig. \ref{fig:upperbounds} we plot the gate errors (dot-dashed lines) to implement the controlled phase gate $\text{CZ}(\pi)$ for different cross-Kerr phase shifts $\chi t = \frac{\pi}{2}, \frac{\pi}{7}, \frac{\pi}{10}$ that are amplified by the factor $\lambda_1$ = 2 (blue squares), $7$ (red circles) and $10$ (green triangles), respectively. The corresponding upper bounds are shown as solid lines.

\section{Photon loss model}  \label{sec: photon loss model}
We model photon losses by coupling each mode via a beam splitter interaction to auxiliary modes $c$ and $d$. The total Hamiltonian including the cross-Kerr interaction is given by  
\begin{equation}
H =\chi a^{\dagger}ab^{\dagger}b + g\left(ac^{\dagger} +  a^{\dagger}c\right) +  g\left(bd^{\dagger} +  b^{\dagger}d\right),
\end{equation}
where $c,d\, (c^{\dagger},d^{\dagger})$ are the annihilation and creation operators of the auxiliary modes and $g$ is the coupling strength. If we assume that the auxiliary modes are in the vacuum state $\ket{00}$, after each time step $\Delta t=\frac{t}{4N}$ of the amplification sequence given in \eqref{eq:two mode HA sequence} the dynamics of the two modes $a$ and $b$ is given by the quantum channel,
\begin{equation}
\label{eq:channelj}
 \mathcal{E}^{(j)}_{\Delta t} (\cdot) = \text{Tr}_{cd} \left[  e^{-iH_{j}\Delta t} \left((\cdot) \otimes \ket{00}\bra{00} \right) e^{iH_{j}\Delta t}   \right],  
\end{equation}
where $\text{Tr}_{cd}$ denotes the partial trace over the auxiliary modes and
\begin{gather}\label{eq:squeezing blocks}
H_{j} = S^{\dagger}_{j} H S_{j},\\ S_{j} \in \left \{ S_{a,\alpha}S_{b,\beta} : j \in \{1,2,3,4\} \, ,\alpha = \pi\, \left \lfloor \frac{j-1}{2} \right \rfloor, \beta = \pi \left(j-1\right)\text{mod}\,2    \right \}.\notag
\end{gather}
If we introduce the loss rate $\eta t$ via choosing $g^{2}\Delta t^{2}= \eta \Delta t$, we obtain for $j = 1,2 $ the quantum channels
\begin{equation}
\begin{aligned}
\mathcal{E}^{(j)}_{\Delta t} = &\text{id} +  \Delta t \,\mathcal H_{j} + {{\Delta t}}\Bigl( \cosh^{2}(r)\mathcal{L}_{l} +  \sinh^{2}(r) \mathcal{L}_{h} - \frac{\sinh(2r)}{2} \left( \mathcal{K}_{a} \pm  \mathcal{K}_{b} \right)  \Bigr) + \mathcal{O}\left(\frac{1}{N^{2}} \right),
\end{aligned}
\end{equation}
where $\mathcal{H}_{j} = -i[H_{j},\cdot] $ , $\mathcal{L}_{l}=\eta(\mathcal D_{a}+\mathcal D_{b})$ and $\mathcal{L}_{h}=\eta(\mathcal D_{a^{\dagger}}+\mathcal D_{b^{\dagger}})$  are  formed by Lindblad operators of the form $\mathcal{D}_{L}$, given in the main body of the manuscript, and we defined the super operators \begin{equation}
    \mathcal{K}_{L}(\cdot) =\eta\left(
     L (\cdot) L + \frac{1}{2} \left( L^{2}(\cdot)+ (\cdot) L^{2} \right)+ L^{\dagger} (\cdot) L^{\dagger} + \frac{1}{2} \left( L^{\dagger 2}(\cdot)+ (\cdot) L^{\dagger 2} \right) \right).
\end{equation}
Similarly, for $j = 3,4 $ we obtain 
\begin{equation}
\begin{aligned}
\mathcal{E}^{(j)}_{\Delta t} = &\text{id} +  \Delta t \,\mathcal H_{j} + {{\Delta t}}\Bigl( \cosh^{2}(r)\mathcal{L}_{l} +  \sinh^{2}(r) \mathcal{L}_{h} + \frac{\sinh(2r)}{2} \left( \mathcal{K}_{a} \mp  \mathcal{K}_{b} \right)  \Bigr) + \mathcal{O}\left(\frac{1}{N^{2}} \right).
\end{aligned}
\end{equation}
As such, the sequence of quantum channels $\left(\prod^{4}_{j =1}  \mathcal{E}_{\Delta t}^{(j)}\right)^{N}$ is in the limit of infinitely many Trotter steps $N$ given by
 \begin{equation}
 \begin{aligned}
 \lim_{N\to\infty}  \left( \prod^{4}_{j =1}  \mathcal{E}^{(j)}_{\Delta t}\right)^{N} 
= & \,  \exp\biggl( \frac{t}{4}  \sum^{4}_{j=1}\left(\mathcal{H}_{j}\right)  +  t \cosh^{2}(r)\mathcal{L}_{l}+ t\sinh^{2}(r) \mathcal{L}_{h} \biggr)\\
= & \,  \exp\biggl(  \left[\mathcal{H}_{\lambda_2}  +   \cosh^{2}(r)\mathcal{L}_{l}+ \sinh^{2}(r) \mathcal{L}_{h}\right] t \biggr).
\end{aligned}
\end{equation}
\begin{figure}[!ht]
 \centering
\includegraphics[scale=0.77]{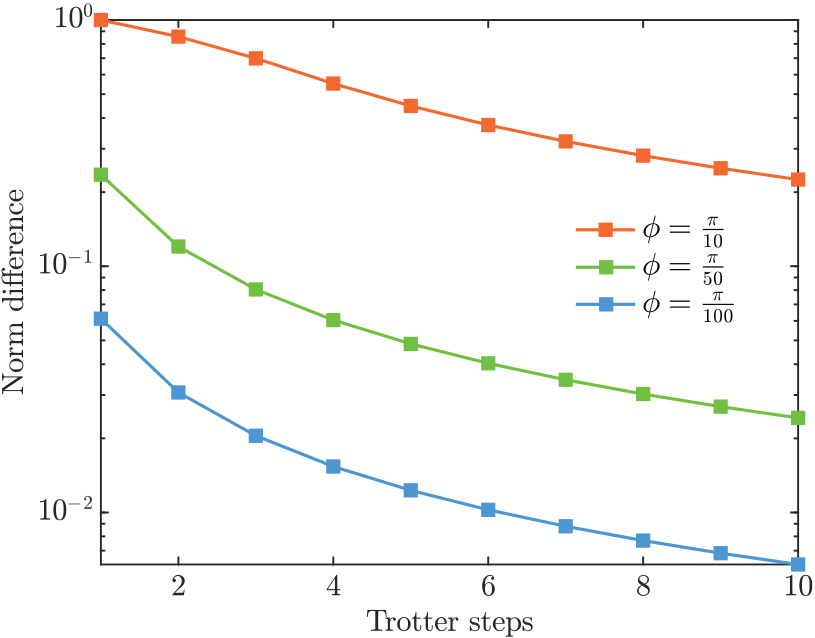}
\caption{\label{fig:normofquantumchannels}Norm difference \eqref{eq:errorChannels} for different phase shifts as a function of the number of Trotter steps $N$. We used $\chi t = 1.2 \times 10^{-3} $ for the bare phase shift and $\eta t = 5.76\times10^{-4}\,\text{dB}$ for the loss rate \cite{venkataraman2013phase, amrani2021low}. For each curve, the norm difference was normalized to $1$ by dividing by its largest value.}
\end{figure}
\\
In Fig. \ref{fig:normofquantumchannels} we investigate the dynamics for finite $N$ by analyzing the error 
\begin{align}
\label{eq:errorChannels}
\epsilon=\left\Vert \mathcal{E}_{t} - \left(\prod_{j=1}^{4} \mathcal E_{\Delta t}^{(j)}\right)^{N}\right\Vert_{\text{HS}}
\end{align}
between the quantum channel $\mathcal E_{t}$, obtained for $N\to \infty$, and the sequence of quantum channels $\left(\prod_{j=1}^{4} \mathcal E_{\Delta t}^{(j)}\right)^{N}$ where the quantum channels $\mathcal E_{\Delta t}^{(j)}$ are given in \eqref{eq:channelj}. We study the error \eqref{eq:errorChannels} between the two evolutions for different phase shifts $\phi=\frac{\pi}{100},\frac{\pi}{50},\frac{\pi}{10}$ obtained by amplifying the bare phase shift $\chi t = 1.2 \times 10^{-3} $  for a loss rate $\eta t = 5.76\times10^{-4}\,\text{dB}$ \cite{venkataraman2013phase, amrani2021low} for different number of Trotters steps $N$.

\section{Fluctuations in the squeezing angles} 
Here, we analyze the performance of the squeezing sequences when errors in the squeezing angles $\theta\in\{0,\pi\}$ are present. We model errors in the squeezing angles $\theta_{a}$ and $\theta_{b}$ for each mode $a$ and $b$ by considering small fluctuations $\theta_{a}\rightarrow \theta_{a}+\delta_{a},~\theta_{b}\rightarrow \theta_{b}+\delta_{b}$ where $\delta_{a}$ and $\delta_{b}$ are drawn from a Gaussian distribution centered around $\mu=0$ and with standard deviation $\sigma$. The squeezing angle perturbed transformations for each mode $a$ and $b$ are given by 
\begin{align}\label{eq:perturbedsqueezing}
S^{(\delta_{a})}_{a,\theta_{a}}=&\exp\left[\frac{r}  {2}  \left(a^{2}e^{-i(\theta_{a} + \delta_{a})}-a^{\dagger 2}e^{i(\theta_{a} + \delta_{a})}\right)   \right],~~~~ S^{(\delta_{b})}_{b,\theta_{b}}=&\exp\left[\frac{r}  {2}  \left(b^{2}e^{-i(\theta_{b} + \delta_{b})}-b^{\dagger 2}e^{i(\theta_{b} + \delta_{b})}\right)   \right].
\end{align}
We assume that the errors $\delta_{a}$ and $\delta_{b}$ are the same for squeezing $S$ and anti squeezing $S^{\dagger}$, i.e., modeling squeezing angle errors by considering blocks of the form $S^{\dagger (\delta_a)}_{a,\theta_{a}}S_{b,\theta_{b}}^{\dagger (\delta_b)}HS^{(\delta_a)}_{a,\theta_{a}}S^{(\delta_b)}_{b,\theta_{b}}$ in the squeezing sequences.

\begin{figure}[!ht]
 \centering
\includegraphics[scale=0.72]{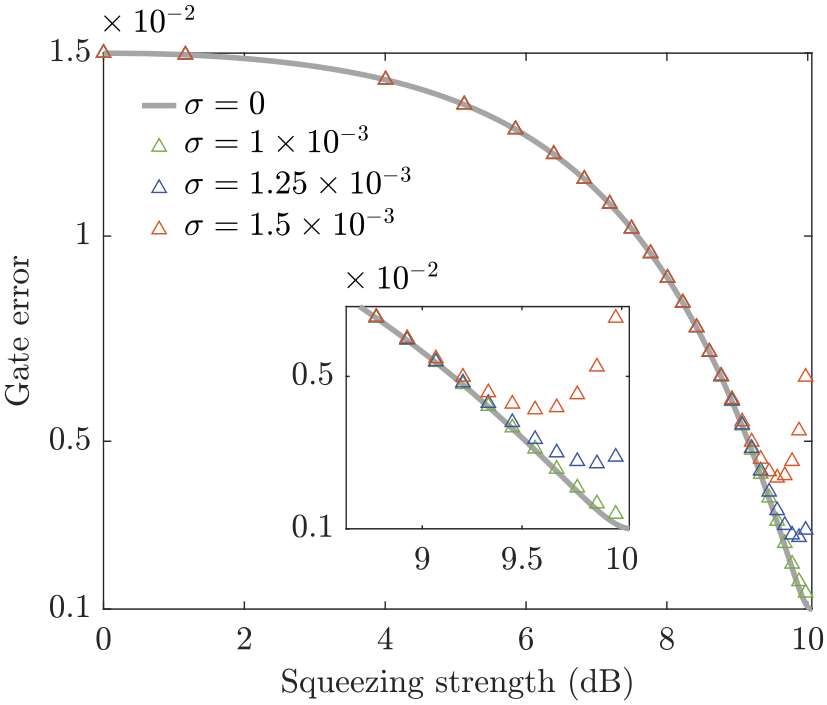}
\caption{\label{fig:errorsinsqueezingangles}Gate error for implementing $\text{CZ}(\frac{\pi}{100})$ gate in the presence of photon losses when the cross-Kerr interaction is amplified through single-mode squeezing transformations that contain errors in the squeezing angle \eqref{eq:perturbedsqueezing} for the same parameter setting as in Fig. 2 (b), right panel. The inset plot shows the gate error for a smaller range of the squeezing strength. The errors $\delta_a, \delta_b$ are drawn from a Gaussian distribution with mean $\mu = 0$ and standard deviations $\sigma = 0$ (grey line, see Fig. 2 (b), right panel), $1\times10^{-3}$ (green triangles), $1.25\times10^{-3}$ (blue triangles) and $1.5\times10^{-3}$ (red triangles). The data points show the average over $15$ samples.}
\end{figure}

In Fig. \ref{fig:errorsinsqueezingangles} we numerically investigate the effect of squeezing angle errors based on the considerations above. We study the same parameter setting as in Fig. 2 (b), right panel, in the main body of the manuscript for different standard deviations $\sigma$.

\end{document}